\documentstyle[aps,preprint]{revtex}
\begin{document}
\draft
\title{\bf Exact, convergent periodic-orbit expansions of
           individual energy eigenvalues of regular quantum graphs}
\author{R. Bl\"umel, Y. Dabaghian and R. V. Jensen}
\address{Department of Physics, Wesleyan University,
Middletown, CT 06459-0155}
\date{\today}
\maketitle

\begin{abstract}
We present exact, explicit, convergent periodic-orbit
expansions for individual
energy levels of regular quantum graphs.
One simple application is the energy levels of a particle
in a piecewise constant potential. Since the classical
ray trajectories (including ray splitting) in such
systems are strongly chaotic, this result provides
the first explicit quantization of a classically
chaotic system.
\end{abstract}


\section{introduction}
Within the framework of semiclassical periodic-orbit theory
the quantization procedures for integrable and chaotic systems
differ substantially. An integrable system
may be quantized using
EBK theory \cite{Gutzw}.
The set of integrals
\begin{equation}
I_{i}=\int_{C_{i}}p_{\mu}d q^{\mu}=h (n_{i}+\mu_{i}),\ \ \
i=1,...,N,
\label{ebk}
\end{equation}
extended along the $N$ fundamental cycles $C_i$
of the $N$-dimensional phase-space tori
yield the (semiclassical)
quantization conditions for every action variable $I_{i}$.
Here the $n_i$'s are integer
quantum numbers and the $\mu_i$'s are Maslov
indices.
Although not exact in general, the quantization condition
(\ref{ebk}) does (implicitly)
produce individual energy levels
$E_{n_1,\ldots,n_N}$ that can be labelled, one by one,
with the $N$ quantum numbers $n_1,\ldots,n_N$.
This procedure differs
markedly from the chaotic case where the focus
is not on
individual energy levels but on {\it global}
characteristics of the spectrum. For instance,
instead of finding individual energy levels as in
(\ref{ebk}), periodic-orbit quantization schemes
for chaotic systems, such as Gutzwiller's trace formula
\cite{Gutzw} compute the density of states
\begin{equation}
\rho (k)= \sum_{j=1}^{\infty}\delta\left(k-k_{j}\right) ,
\label{density}
\end{equation}
from which individual energy levels are extracted
indirectly as the singularities of $\rho$.
In a chaotic system
the only available classical input are the periodic orbits of
the system and the density of states (\ref{density})
is computed according to \cite{Gutzw}
\begin{equation}
\rho (k)\approx\bar{\rho}(k)+\frac{1}{\pi}\mathop{\rm Im}
\sum_{p}T_{p}(E)\sum_{\nu=1}^{\infty}
A_{p}^{\nu}(E)\,e^{i\nu S_{p}(E)}.
\label{rho}
\end{equation}
Here $\bar{\rho}(k)$ is the average density of states,
$S_{p}(E)$, $T_{p}(E)$ and $A_{p}(E)$ are correspondingly
the action, the period and the weight factor of the prime
periodic orbit labeled by $p$, and $\nu$ is the repetition index.
Again, the scheme (\ref{rho}) is not usually exact. More
seriously, however, in contrast to (\ref{ebk}) it fails
to produce individual energy levels in the form
``$E_n=\ldots$''.
The difference between (\ref{ebk}) and (\ref{rho}) cannot
be emphasized enough. While (\ref{ebk}) allows us to
``pick and choose'' a particular energy eigenvalue,
in the chaotic case
all of the eigenvalues have to be computed according to
(\ref{rho}), and only a subsequent non-analytic inspection
and counting procedure allows us to focus on an individual
energy level.
There is, however, a class of quantum chaotic
systems, regular quantum graphs \cite{Opus}, which
are explicitly solvable analytically \cite{Opus},
i.e. exact periodic-orbit
expansions
of the form ``$E_n=\ldots$'' exist.
The purpose of this paper is to expand
considerably with respect to the work presented
in \cite{Opus} and to present a thorough discussion
of our methods and their validity.

The organization of this paper is as follows. In Sect. II
we extend the theory of
quantum graphs \cite{QGT1,QGT2,QGT3,QGT4,QGT5} to include
dressed graphs, i.e.
quantum graphs with
arbitrary potentials on their bonds.
In Sect. III
we define regular quantum graphs and present explicit,
convergent periodic-orbit
expansions of individual
eigenvalues. These expansions are not just formal
identities; the periodic-orbit
expansions presented in Sect. III converge, and
converge to the correct eigenvalues.
In Sect. IV we present a worked example of a simple
quantum graph whose spectrum is computed in three
different ways: (i) numerically exactly,
(ii) via the explicit periodic-orbit expansions
presented in
Sect. III and (iii) via numerical integration
using an exact trace formula for the density
of states. The results of the three methods
agree. This proves validity and convergence
of our approach.
In Sect. V we summarize our results
and conclude the paper.

\section{Dressed Quantum Graphs}
A quantum graph consists of a
quantum particle whose motion is confined
to a one-dimensional
network of $N_{B}$ bonds $B_{ij}$
connecting $N_{V}$ vertices
$V_{i}$. An example of a graph with six vertices and ten
bonds is shown in Fig.~1.
The topology of a given graph is fully characterized by its
connectivity matrix $C_{ij}$:
\begin{equation}
C_{ij}=C_{ji}=
\cases{ 1, &if $V_{i}$ and $V_{j}$
are connected, \cr 0, &if they are not.\cr}
\label{con}
\end{equation}
Every bond $B_{ij}$ which connects the vertices $V_{i}$ and $V_{j}$,
supports a solution $\psi_{ij}(x)$ of the Schr\"odinger equation
\begin{equation}
\left(-i\frac{d}{dx}-A_{ij}\right)^{2}\psi_{ij}(x)=E\psi_{ij}(x).
\label{sch}
\end{equation}
Here $0\leq x\leq L_{ij}$ is the coordinate along $B_{ij}$ measured
from $V_i$ to $V_j$, and $L_{ij}=L_{ji}$ is the length of the bond.
A constant, real, skew symmetric matrix $A_{ij}=-A_{ji}$,
which plays the role of a magnetic field vector potential,
is sometimes introduced as a tool for braking the time-reversal
symmetry, which, in turn, is known to affect the statistics of
the level distribution \cite{BGS,LH}.

In this paper we generalize
the Schr\"odinger operator in (\ref{sch})
by adding potentials $U_{ij}(x,E)$ on the graph bonds. We call this
generalization ``dressing of the graph''.
While in general the potentials $U_{ij}(x,E)$ may depend on the bond
coordinate $x$ and the energy $E$ in an arbitrary way, we restrict
ourselves here to the {\em scaling} case
\begin{equation}
U_{ij}(E)=\lambda_{ij}E, \ \ \lambda_{ij}=\lambda_{ji},
\label{pot}
\end{equation}
which allows us to introduce physical parallels
between quantum graphs and ray-splitting systems
\cite{Couch,RS1,RS2}.
A quantum graph with the potentials (\ref{pot}) on its bonds can also
be viewed as a generalized step potential, such as the one
shown in Fig.~2~(a). These potentials were studied before in great
detail in connection with Anderson localization \cite{Pastur}.
Potentials of this type can be represented by a linear graph,
such as the
one shown in Fig.~2~(b).
Scaled potentials such as (\ref{pot})
cast the Schr\"{o}dinger equation
into the form
\begin{equation}
\left(-i\frac{d}{dx}-A_{ij}\right)^{2}\psi_{ij}(x)=
\beta_{ij}^2 E\psi_{ij}(x),
\label{schred}
\end{equation}
where the parameters $\beta_{ij}^2=1-\lambda_{ij}$,
$\beta_{ij}=\beta_{ji}$ are defined on the
corresponding bonds $B_{ij}$.

Depending on whether the energy $E=k^{2}$ of the particle is above
or below the scaled potential height $U_{ij}(E)$, the solution of
equation (\ref{schred}) on the bond $B_{ij}$ is either a combination
of the free waves,
\begin{equation}
\psi_{ij}(x)=
a_{ij}\frac{e^{i\left(-\beta _{ij}k+A_{ij}\right)x}}
{\sqrt{\beta_{ij}k}}
+b_{ij}\frac{e^{i\left(\beta _{ij}k+A_{ij}\right)x}}
{\sqrt{\beta_{ij}k}},
\ \ \lambda_{ij}<1,
\label{psi}
\end{equation}
or a combination of the tunneling solutions,
\begin{equation}
\psi_{ij}(x)=a_{ij}e^{\left( -\beta _{ij}k+iA_{ij}\right)x}
 +b_{ij}e^{\left( \beta _{ij}k+iA_{ij}\right) x}, \ \ \lambda_{ij}>1,
\label{tunnel}
\end{equation}
where the factors $(\beta_{ij}k)^{-1/2}$ in the propagating waves
(\ref{psi}) were introduced to insure
proper flux normalization \cite{LL}.
Due to the scaling assumption, there is no transition between these
two cases as a function of $E$. From now
on we shall assume that the energy $E$ is kept above the
maximal scaled potential height,
\begin{equation}
\lambda_{ij}<1, \ \ i,j=1,...,N_{V}, \ \ C_{ij}\neq 0,
\label{above}
\end{equation}
which will allow us to exclude the tunneling solutions (\ref{tunnel}).
At every vertex $V_{i}$, the bond wave functions satisfy the
continuity conditions
\begin{equation}
\psi_{ij}(x)\mid_{x=0}=\varphi_{i}C_{ij}, \ \ i,j=1,...,N_{V}
\label{phase1}
\end{equation}
and the current conservation conditions
\begin{equation}
\sum_{j=1}^{N_{V}}C_{ij}\left(-i\frac{d}{dx}-A_{ij}\right)\psi_{ij}(x)
\mid_{x=0}= -i\lambda_{i}\varphi_{i},\ \ \  i,j=1,...,N_{V}.
\label{phase2}
\end{equation}
Here $\varphi_{i}$ is the value of the wave
function at the vertex $V_{i}$,
and the $\lambda_{i}$'s are free parameters of the problem for which
the scaling is introduced as
\begin{equation}
\lambda_{i}=\lambda^{0}_{i}k.
\label{sca}
\end{equation}
The conditions
(\ref{phase1}) and (\ref{phase2}) are consistent only for a
discrete set of wave numbers $k_n $, which defines the
spectrum of the quantum graph problem.
Since $\psi_{ij}(x)$ and $\psi_{ji}(y)$ represent the same
wave function on the bond connecting the vertices $V_i$ and
$V_j$ (the only difference is that $x$ is measured from
vertex $V_i$ and $y$ is measured from vertex $V_j$) we have
\begin{equation}
\psi_{ji}\left(L_{ij}-x\right) =\psi_{ij}(x).
\label{symm}
\end{equation}
Using (\ref{psi}) we obtain
\begin{equation}
\psi_{ji}\left( L_{ij}-x\right) =
a_{ji}\frac{e^{i\left(-\beta_{ij}k+A_{ji}\right)
\left(L_{ij}-x\right)}}{\sqrt{\beta_{ij}k}} +
b_{ji}\frac{e^{i\left( \beta_{ij}k+A_{ji}\right)
\left(L_{ij}-x\right)}}{\sqrt{\beta_{ij}k}}
=\psi_{ij}(x).
\end{equation}
Therefore
the coefficients $a_{ij}$ and $b_{ij}$ are related according to
\begin{eqnarray}
a_{ji}=b_{ij}e^{i\left( \beta _{ij}k+A_{ij}\right) L_{ij}},\ \ \
b_{ji}=a_{ij}e^{i\left( -\beta _{ij}k+A_{ij}\right) L_{ij}}.
\label{ab}
\end{eqnarray}
The coefficients $a_{ij}$ and $a_{ji}$, ($b_{ij}$ and $b_{ji}$) are
considered different. This implies that
the bonds of the graph are directed.
Equations (\ref{ab}) can be written in matrix form,
\begin{equation}
\vec{a}=P\tilde D(k)\vec{b},
\label{Dmat}
\end{equation}
where $\vec{a}$ and $\vec{b}$ are the
$2N_{B}$-dimensional vectors of
coefficients, $\tilde D$ is a diagonal matrix in the
$2N_{B}\times 2N_{B}$ space of directed bonds and
\begin{equation}
P=\pmatrix{0 & 1_{N_{B}} \cr
 1_{N_{B}} & 0},
\label{P}
\end{equation}
where $1_{N_{B}}$ is the $N_{B}$-dimensional unit matrix.
Explicitly we have
\begin{equation}
\tilde D_{ij,pq}(k)=\delta _{ip}\delta _{jq}
e^{i\left( \beta _{ij}k+A_{ij}\right)
L_{ij}}.
\label{tild}
\end{equation}
The pairs of indices $(ij)$, $(pq)$, identifying the
bonds of the graph
$\Gamma$, play the role of the indices of
the matrix $\tilde D(k)$.
Alternatively the
wave function can be written as a linear combination of plane
waves scattering off $V_{i}$.
An incoming wave with normalized flux on
the bond $B_{j'i}$ gives rise
to a partial wave contribution scattering into bond $B_{ij}$
according to
\begin{equation}
\psi^{(i)}_{jj'}(x_{j})=
\delta_{jj'}\frac{e^{i\left(-\beta_{ij}k+A_{ij}\right)x_{j}}}
{\sqrt{\beta_{ij}k}}
+
\sigma_{j,j'}^{(i)}\frac{e^{i\left(\beta_{ij}k+A_{ij}\right)
x_{j}}}{\sqrt{\beta_{ij}k}}.
\label{scatt}
\end{equation}
Here $\sigma _{j,j'}^{(i)}(k)$ is the matrix element of the
vertex scattering matrix $\sigma^{(i)}(k)$, which distributes
the incoming flux on bond $B_{j'i}$ into the bond $B_{ij}$.
The wave function $\psi_{ij}(x_{j})$ on the bond $B_{ij}$ is a
superposition of the partial waves (\ref{scatt}) with amplitudes
$a_{ij'}$ corresponding to the incoming flux on the bond
$B_{j'i}$ towards the vertex $V_{i}$, i.e.
\begin{equation}
\psi_{ij}(x_{j})=\sum_{j'}a_{ij'}\psi_{jj'}^{(i)}(x_{j}).
\label{pssi}
\end{equation}
Using the representation (\ref{psi}) of $\psi_{ij}$ in (\ref{pssi})
and comparing coefficients yields
\begin{equation}
b_{ij}=\sum_{j'}\sigma_{j,j'}^{(i)}a_{ij'}.
\label{ab1}
\end{equation}
Substituting (\ref{scatt}) into the boundary conditions, we
obtain the vertex scattering matrix
\begin{equation}
\sigma_{j,j'}^{(i)}\equiv\sigma_{ji,ij'}^{(i)}=\left(-\delta_{jj'}+
\frac{2\sqrt{\beta_{ij}\beta_{ij'}}}{v_{i}+i\lambda^{0}_{i}}\right)
C_{ji}C_{ij'},
\label{sigma}
\end{equation}
with
\begin{equation}
v_i=\sum_{j=1}^{N_{V}}\beta_{ij}C_{ij}.
\label{vi}
\end{equation}
We see that
in the scaling case the matrix elements $\sigma _{j,j'}^{(i)}$ of
the vertex scattering matrix $\sigma^{(i)}$ are $k$-independent
constants.
The matrix element $\sigma _{j,j}^{(i)}$
has the meaning of the
reflection coefficient from the vertex
$V_{i}$ along the bond $B_{ij}$,
and the elements $\sigma _{j,j'}^{(i)}$,
$j\neq j'$ are the transmission
coefficients for transitions between different bonds.
Equation (\ref{ab1}) can be written as
\begin{equation}
\vec b=\tilde T\vec a,
\label{Tmat}
\end{equation}
where
\begin{equation}
\tilde T\equiv \tilde T_{ij,nm}=\delta _{in}C_{ji}C_{nm}
\sigma_{j,m}^{(i)}.
\label{tilt}
\end{equation}
Equations (\ref{Dmat}) and (\ref{Tmat}) together result in
\begin{equation}
\vec a=S(k)\vec a,
\label{lin}
\end{equation}
where $S(k)$ (the total graph
scattering matrix) is given by
\begin{equation}
S(k)=D(k)T
\label{S}
\end{equation}
and $D=P\tilde D P$ and $T=P\tilde T$.
The consistency of the system of linear equations (\ref{lin})
requires the spectral equation
\begin{equation}
\Delta (k)=\det \left[1-S(k)\right]=0
\label{det}
\end{equation}
to be satisfied. This condition defines
the set of allowed momenta $\{k_n \}$.

The density of the momentum states of
the dressed quantum graph is given by
\begin{equation}
\rho(k)=\sum_{n=1}^{\infty}\delta(k-k_n ),
\label{rho1}
\end{equation}
where the $k_n $'s are the solutions of (\ref{det}).
An exact periodic-orbit expansion
for $\rho(k)$ can be obtained
directly from the spectral
equation (\ref{det}) as follows
\cite{QGT1,QGT2,QGT3,QGT4,QGT5}.
The logarithmic derivative
of (\ref{det}) is singular at
every one of its roots. Between roots, the phase of the
spectral determinant varies slowly such that
\begin{equation}
\rho(k)\ =\ \bar\rho(k)\, -{1\over\pi}\,
\lim_{\epsilon \rightarrow 0}\,
{\rm Im}\,
\frac{d}{dk}
\ln\det\left[1-S\left(k+i\epsilon\right)\right].
\label{exp}
\end{equation}
Using the well-known identity
\cite{QGT1,QGT2,QGT3,QGT4,QGT5}
\begin{equation}
\ln\det[1-S]=-{\rm Tr}\, \sum_{n=1}^{\infty}\, {1\over n}\, S^n,
\end{equation}
we obtain
\begin{equation}
\rho (k)=\bar\rho(k)\, +\,
\frac{1}{\pi}\lim_{\epsilon\rightarrow 0}\, {\rm Im}\,
\frac{d}{dk}\sum_{n=1}^{\infty}
\frac{1}{n}\mathop{\rm Tr}\left[S(k+i\epsilon)\right]^{n}.
\label{exp1}
\end{equation}
Since the matrix indices of (\ref{S})
correspond to the vertices of the
graph, the trace of the {\em n}th power of the scattering matrix
can be interpreted as a sum over all closed connected sequences
consisting of
$n$ bonds
\cite{QGT1,QGT2,QGT3,QGT4,QGT5}.
Classically, these periodic connected
sequences of $n$ bonds $B_{ij}$
correspond to the periodic orbits traced by a point particle
moving on the graph.
Geometry and proliferation properties of the periodic orbits
are determined
completely by
the topology of the graph.

The behavior of the periodic orbits on graphs exhibits the typical
features of chaotic systems.
The meaning of classical chaoticity on graphs is well defined, as
demonstrated in the following.
A classical graph system consists of a graph $\Gamma $ and a point
particle
moving along its bonds, which scatters elastically
at every vertex $V_{i}$ along
the direction of any of the bonds emanating from
this particular vertex, with
different probabilities. The probability amplitudes
for every scattering
channel can be obtained in the short wavelength limit
from the quantum
mechanical transition amplitudes defined at every
vertex $V_{i}$ by the
corresponding scattering matrix $\sigma _{j,j^{\prime}}^{(i)}$.
In the scaling case, the matrix
elements $\sigma _{j,j^{\prime}}^{(i)}$ are
$k$-independent constants and thus do not
depend on $\hbar$ at all.
Therefore the same matrix elements determine both the quantum
and the classical scattering probabilities.

 For every given graph $\Gamma$ the global average
rate of exponential proliferation
of periodic orbits, the topological
entropy $\Lambda$,
is given by
\begin{equation}
\Lambda=\lim_{l\rightarrow\infty}
{\ln\left[\#(l)\right]\over l},
\label{topo}
\end{equation}
where $l$ characterizes the length of the periodic orbits
(for instance their code lengths) and
$\#(l)$ is the total number of periodic orbits
of length $\leq l$.
The number of possible
periodic orbits increases exponentially with their lengths,
(or, equivalently, the number of scattering events)
with a rate $\Lambda$ which
depends only on the topology of the graph.
Since the phase space of the system is bounded,
the dynamics of such a particle is mixing \cite{QGT1}.

Since we are focusing on the case $\lambda_{ij}<1$, most of the
classical periodic orbits on a graph are above-barrier
reflection orbits as illustrated in Figs.~2 and 3.
In the context of ray splitting these orbits are also
known as non-Newtonian orbits \cite{RS3,Bauch,Nova}.
The inclusion of all non-Newtonian orbits in our periodic-orbit
expansions of individual eigenvalues discussed below is
crucial for rendering these expansions exact.

Traversing the bond $B_{ij}$ contributes the amount
\begin{equation}
S_{ij}=\int_{B_{ij}}\left( \beta _{ij}k+A_{ij}\right) dx
\label{action}
\end{equation}
to the total action of the trajectory traced by the particle.
These actions appear in the phases of the
exact wave functions (\ref{psi}).
This connection means that the semiclassical (eikonal) form is
exact for the quantum graph wave functions. More importantly,
the amplitudes $e^{iS_{ij}}$ determine the
matrix $D(k)$, and hence
the scattering matrix $S(k)$.
As a consequence, the ``closed bond sequence expansion''
(\ref{exp1}) can be written explicitly as a periodic-orbit
expansion in terms of the phases
(\ref{action}),
\begin{equation}
\rho (k)=\bar{\rho}(k)+\frac{1}{\pi}\mathop{\rm Re}
\sum_{p}T_{p}(k)
\sum_{\nu=1}^{\infty}\tilde A_{p}^{\nu}\,e^{i\nu
\tilde S_{p}(k)},
\label{r}
\end{equation}
where $\tilde S_{p}$ is the action of the prime periodic orbit $p$
composed of the partial actions $S_{ij}$ of equation (\ref{action})
accumulated along the periodic orbit $p$,
and $T_p(k)=d\tilde S_p(k)/dk$.
The first term in (\ref{r}) describes the
average behavior of the density of states while
the second represents the
fluctuations around the average.
The amplitude of
every periodic orbit $p$ contains the constant factor
$\exp({i\sum_{ij}A_{ij}L_{ij}})$. This factor
can be absorbed into the weight
factor $\tilde A_{p}$.
Thus, defining the reduced classical bond actions
\begin{equation}
S_{ij}^0=\beta_{ij}L_{ij}
\label{redbond}
\end{equation}
and the total reduced action $S_p^0$ accumulated along
the periodic orbit $p$,
\begin{equation}
S_p^0 =\sum_{ij\ {\rm along}\ p}S_{ij}^0 ,
\end{equation}
the final periodic-orbit expansion for the density of states
for scaling systems can
be written as
\begin{equation}
\rho (k)=\bar{\rho}(k)+\frac{1}{\pi}\mathop{\rm Re}
\sum_{p} S_p^0
\sum_{\nu=1}^{\infty}A_{p}^{\nu}\,e^{i\nu S_{p}^0 k}.
\label{ro}
\end{equation}
In contrast with (\ref{rho}), the expression (\ref{ro}) for
the density of states
is exact;
the action lengths $S_{p}^{0}$ and the weight
factors $A_{p}$ are $k$-independent constants.

The staircase function
\begin{equation}
N(k)=\sum_{n=1}^{\infty}\Theta (k-k_n )
\label{N}
\end{equation}
is obtained by direct integration of (\ref{rho1}).
Using (\ref{ro}), $N(k)$ can be expanded as
\begin{equation}
N(k)=\bar N(k)\ -\
\frac{1}{\pi}
\lim_{\epsilon \rightarrow 0}\,
{\rm Im}\,
\ln\det
\left[1-S(k+i\epsilon)\right]=\bar{N}(k)+
{1\over\pi}\, \mathop{\rm Im}
\sum_{p}\sum_{\nu =1}^{\infty}\frac{A_{p}^{\nu}}{\nu}
e^{i\nu S_{p}^{0}k}.
\label{stair}
\end{equation}
Just like (\ref{ro}) this expansion is exact.
The first term represents the average behavior of the staircase;
the second term describes the fluctuations
around the average.

\section{Regular quantum graphs and explicit spectral formula}
Since the scattering matrix (\ref{S})
is a unitary matrix, its eigenvalues
have the form $s_{l}=e^{i\theta _{l}(k)}$. Therefore the spectral
determinant (\ref{det}) can be written as
\begin{eqnarray}
\Delta (k)=\prod_{l=1}^{2N_{B}}\left[1-e^{i\theta_{l}(k)}\right]=
\left[ 1-\sum_{l=1}^{2N_{B}}e^{i\theta_{l}(k)}+...
         +e^{i\sum_{l=1}^{2N_{B}}\theta _{l}(k)}\right]
\cr =
2e^{i\Theta_{0}(k)}\left[\cos[\Theta_0(k)]
+\sum_{j=1}^{N_C-1}(-1)^j\cos [\Theta_j(k)]\right],
\label{expa}
\end{eqnarray}
where
\begin{equation}
\Theta_0={1\over 2}\, \sum_{l=1}^{2N_{B}}\theta_{l}(k)
\label{Theta0}
\end{equation}
is the total phase of the spectral determinant,
the $\Theta_j$'s in (\ref{expa})
are linear
combinations (sums and differences)
of the phases
$\theta_{l}(k)$ and $N_C=4^{N_B}/2$.
Evaluated directly, the spectral determinant is a
polynomial of the (complex) matrix elements (\ref{tild})
with coefficients that are determined by the
matrix elements
(\ref{tilt}). Factoring
out the total phase (\ref{Theta0}) of this polynomial,
we obtain the spectral equation in the form
\begin{equation}
\cos \left(S_0^0 k-\pi \gamma_{0} \right)=
\Phi(k),
\label{eqn1}
\end{equation}
where
\begin{equation}
\Phi(k)=
\sum_{i=1}^{N_{\Gamma}}a_i\cos(\Omega_i k-\pi\gamma_i).
\label{eqn2}
\end{equation}
Here, based on the reduced bond actions defined in
(\ref{redbond}),
$S_0^0=\sum_{ij}C_{ij}S_{ij}^0$ is
the total reduced action length of the graph,
the frequencies $\Omega_i < S_0^0$ are sums and differences
of the
reduced bond actions
$S_{ij}^{0}$ and $\gamma_{0}$,
$\gamma_{i}$ are constants.
For a general graph $\Gamma$ it is difficult to calculate the
precise number of $\cos$-terms
$N_{\Gamma}$ in (\ref{eqn2}).
But an upper limit is given by the number of
possible linear combinations of the $N_B$ reduced bond
actions $S_{ij}^0$.
Since there are $2^j\left(\matrix{N_B\cr j\cr}\right)$
ways of picking $j$ actions out of $N_B$ possible ones
and combining them with ``$+$'' and ``$-$'' signs, we
obtain
\begin{equation}
N_{\Gamma}\leq \sum_{j=1}^{N_B}\, 2^j\,
\left(\matrix{N_B\cr j\cr}\right)=
-1+
 \sum_{j=0}^{N_B}\,
\left(\matrix{N_B\cr j\cr}\right)2^j\, 1^{N_B-j}=
3^{N_B}-1.
\label{estimate}
\end{equation}

A graph $\Gamma$ is called {\it regular}, if the
condition
\begin{equation}
\sum_{i=1}^{N_{\Gamma}}\, |a_i|=\alpha <1
\label{reg}
\end{equation}
is fulfilled.
In case the condition
(\ref{reg}) is satisfied,
the spectral
equation (\ref{eqn1}) can be immediately solved
to yield the following
implicit equation for
the eigenvalues
\begin{equation}
k_n ={\pi\over S_0^0}\left[n+\mu+\gamma_{0}\right] + {1\over S_0^0}
\cases{\arccos[\Phi(k_n)], &for $n+\mu$ even \cr
        \pi-\arccos[\Phi(k_n)], &for $n+\mu$ odd, \cr}
\label{levels}
\end{equation}
where $\mu$ is a fixed integer, chosen such that $k_1$ is
the first positive solution of (\ref{eqn1}). Equation
(\ref{levels}) implies the existence of {\it separating points}
\begin{equation}
\hat k_n =\frac{\pi}{S_0^0}(n+\mu+\gamma_0+1)
\label{mid}
\end{equation}
in the spectrum $k_n$ of (\ref{eqn1}). Because of
(\ref{reg})
the points $\hat k_n$
are never solutions of (\ref{levels}). They act as separators
between $k_n$ and $k_{n+1}$.
Since the second term in (\ref{levels})
is bounded by $\pi/S_0^0$, the
deviation $\mid k_n-\hat k_n\mid$
never exceeds $\pi /S_0^0$ for any $n$.
We emphasize, that the separators $\hat k_n $
do not coincide with the average
values $\bar k_n $ of the roots $k_n $.
The explicit decomposition of the
roots $k_n $ into an average part
$\bar k_n $ and a fluctuating part $\tilde k_n $,
$k_n =\bar k_n +\tilde k_n $, can be obtained from
the following
equivalent formulation of (\ref{levels}),
\begin{equation}
k_n ={\pi\over S_0^0}\left[n+\mu+\gamma_{0}+{1\over 2}\right] +
{(-1)^{n+\mu}\over S_0^0}
\left\{\arccos[\Phi(k_n)]-\frac{\pi}{2}\right\}.
\label{average}
\end{equation}
This form of $k_n$ together with the boundedness
of the second term in (\ref{average}) proves
rigorously that $\bar N(k)$, $\bar\rho(k)$ are of the form
\begin{equation}
\bar N(k)={S_0^0\over\pi}k+\bar N(0),\ \ \ \bar\rho(k) =
 {d\bar N(k)\over dk}= {S_0^0\over\pi}.
\label{bnbr}
\end{equation}
Since $\Phi(k)$ contains only frequencies smaller than
$S_0^0$, every open
interval $I_{n}=(\hat k_{n-1},\hat k_n )$
contains one and only one root, i.e. $k_n$.
Moreover, if (\ref{reg}) is fulfilled,
the allowed zones $Z_{n}\subset I_{n}$
where the roots $k_n $ can be
found narrows to
\begin{equation}
k_n \in Z_{n}\equiv\left(\frac{\pi}{S_0^0}
\left(n+\mu+\gamma_0+u\right) ,
\frac{\pi}{S_0^0}\left(n+\mu+\gamma_0+1-u\right)\right),
\label{int}
\end{equation}
where $u=\arccos(\alpha)/S_0^0$.
Correspondingly, there exist forbidden regions $R_{n}$,
\begin{equation}
R_{n}\equiv\left(\frac{\pi}{S_0^0}
\left(n+\mu+\gamma_0+1-u\right) ,
\frac{\pi}{S_0^0}\left(n+\mu+\gamma_0+1+u\right)\right),
\label{zap}
\end{equation}
where roots of (\ref{levels}) can never be found.
Note that $\hat k_n\in R_n$.
In the limit $\alpha\rightarrow 1$
($u\rightarrow 0$), the width
of the forbidden region $R_n$
shrinks to zero,
and the allowed zone $Z_n$ occupies the
whole root interval $I_n$.

The existence of the separating points (\ref{mid}),
is the key to
obtaining the explicit form of the
periodic-orbit expansion for individual roots $k_n$.
Multiplying both sides of (\ref{ro}) by $k$
and integrating from $\hat k_{n-1}$ to $\hat k_n$
we obtain
\begin{eqnarray}
k_n = \hat k_n - {\pi\over 2 S_0^0} -
{1\over\pi}{\rm Re}\sum_p \sum_{\nu=1}^{\infty}A_p^{\nu}\,
{e^{i\nu S_p^0\hat k_n}\over \nu } \left\{
(1-e^{-i\nu\omega_p})\left(i\hat k_n-{1\over\nu S_p^0}\right)
+{i\pi\over S_0^0}e^{-i\nu\omega_p}\right\},
\label{kn}
\end{eqnarray}
where we used (\ref{bnbr}) for the integral over $k\bar\rho$
and defined
$\omega_{p}=\pi S_{p}^{0}/S_0^0$.
Since all the quantities on the right-hand side of (\ref{kn})
are known,
this formula provides an explicit
representation of the roots $k_n$ of the
spectral equation (\ref{det}) in
terms of the geometric characteristics and
the classical properties of the graph.

In \cite{JMP} a mathematical proof is presented that
assures us that
(\ref{kn}) converges. In addition it is proved in
\cite{JMP} that (\ref{kn}) converges to the exact
spectral eigenvalues.
Both convergence and convergence to
the correct results are illustrated with the help
of a specific example in Sect. IV.
It is also proved in \cite{JMP} that the series
(\ref{kn}) is only conditionally convergent.
This means that
for proper convergence the ordering of the terms in (\ref{kn})
is important. Proper
convergence of (\ref{kn}) is obtained if the terms
in (\ref{kn}) are ordered according to the code lengths of the
periodic orbits \cite{JMP}.
In other words, the sum in (\ref{kn}) is to
be extended over all periodic orbits with
code lengths smaller than or equal to $l$
which yields the approximation
$k_n(l)$ to $k_n$. Then, on the basis of the results obtained
in \cite{JMP}, we have $\lim_{l\rightarrow\infty}\, k_n(l)=k_n$.
This means that (\ref{kn}) is exact.
It is important to note here that the ordering of terms in
(\ref{kn}) is {\it not} according to their action lengths, but
according to the lengths of the code words that code for
the periodic orbits. This is intuitively understandable, since
the code length $l$ is connected to the
power $n$ of the $S$-matrix in (\ref{exp1}) according to
$l=n/2$.

The expansion (\ref{kn})
provides an explicit representation of the roots of the
spectral equation (\ref{det}) in terms
of the geometric characteristics
of the graph.
In a similar way one can obtain explicit
expansions for any power of
the energy levels, $k_n ^{m}$, or any
function of the eigenvalues
$f\left(k_n \right)$.

\section{Examples}
The coefficients $A_p$ in (\ref{kn})
assume a particularly simple form in the case of
linear graphs with zero sources,
$\lambda^{0}_{i}=0$, $i=1,...,N_{V}$.
Both the vertices and the bonds of
a linear graph can be naturally
labeled by means of a single index such that
$B_{1,2}\equiv B_{1}$, $B_{2,3}
\equiv B_{2}$,...,$B_{N_V-1,N_V}\equiv B_{N_V-1}$
(see Fig.~2).
The scaling coefficients for the
momentum of the particle are
correspondingly
$\beta_{1,2}\equiv \beta_{1}$,
$\beta_{2,3}\equiv \beta_{2}$,...,$
\beta_{N_V-1,N_V}\equiv \beta_{N_V-1}$,
the bond lengths are
$L_{1,2}\equiv L_{1}$,
$L_{2,3}\equiv L_{2}$,...,$
L_{N_V-1,N_V}\equiv L_{N_V-1}$,
the potentials are
$U_{1,2}\equiv U_{1}$,
$U_{2,3}\equiv U_{2}$,...,$
U_{N_V-1,N_V}\equiv U_{N_V-1}$
and the reduced bond actions are
$S_{1,2}^0\equiv S_{1}^0=\beta_1 L_1$,
$S_{2,3}^0\equiv S_{2}^0=\beta_2 L_2$,...,$
S_{N_V-1,N_V}^0\equiv S_{N_V-1}^0=\beta_{N_V-1} L_{N_V-1}$,
respectively.
In this case, if a prime
periodic orbit $p$ undergoes
$\sigma_{p}^{i}$ reflections from a vertex $V_{i}$ and
$2\tau_{p}^{i}$ transmissions through it, the weight
coefficient in the expansion (\ref{ro}) is \cite{Nova}
\begin{eqnarray}
A_{p}=\prod_{i}r_{i}^{\sigma_{p}^{i}}
(1-r^{2}_{i})^{\tau_{p}^{i}},
\label{A}
\end{eqnarray}
where $r_{i}$ is the reflection
coefficient from the vertex $V_{i}$,
and the product is taken over all the vertices encountered
by the orbit $p$.
If a particle reflects from the
vertex $V_{i}$ traveling along the
bond $B_{i}$, the reflection coefficient is
\begin{eqnarray}
r_i={\beta_{i-1}-\beta_i\over \beta_{i-1}+\beta_i}, \ \ \
i=2,\ldots,N_V-1,\ \ \
r_1=-1,\ \ \ r_{N_V}=-1.
\label{ref}
\end{eqnarray}
We assumed Dirichlet boundary conditions at the left and right
dead ends of the graph.
The reflection coefficient changes its sign
if the reflection happens from the side of the bond
$B_{i+1}$.
If, for a given orbit, the total number of reflections with
$r_{i}<0$ is $\chi_{p}$, then
\begin{eqnarray}
A_{p}=(-1)^{\chi_{p}}\prod_{i}|r_{i}|^{\sigma_{p}^{i}}
(1-r^{2}_{i})^{\tau_{p}^{i}}.
\label{weight1}
\end{eqnarray}
The two-vertex linear graph is trivial and corresponds
to a quantum particle in a square-well box.
A quantum particle moving in a
scaling step potential as shown in Fig.~3,
gives rise to
the simplest non-trivial graph, the scaling
three-vertex linear graph, shown on the bottom of
of Fig.~3. In this case there
is only one non-trivial reflection coefficient,
\begin{eqnarray}
r_2={\beta_1-\beta_2\over \beta_1+\beta_2}.
\label{ref1}
\end{eqnarray}
All the periodic orbits of the
three-vertex linear graph shown in Fig.~3
correspond one to one
with words formed
from a binary code with two letters $\cal L$
and $\cal R$ \cite{Nova,JMP,Forma},
where $\cal L$ stands for a reflection of the orbit off the
left-most vertex
(left-hand potential wall), and $\cal R$
stands for a reflection off the
right-most vertex (right-hand wall).
Thus the
$\cal L$, $\cal R$ code is unique and complete.
For this system the spectral equation is
\begin{equation}
\sin (S_0^0 k)-r_2\sin(\Omega_1 k) =0,
\label{3hydra}
\end{equation}
where $S_0^0=S_1^0+S_2^0$ is the total reduced action
of the graph and
$\Omega_1=S_1^0-S_2^0$. With
$a_1=r_2$ and
$\gamma_0=\gamma_1=\pi/2$, (\ref{3hydra}) is of the
form (\ref{eqn1}), (\ref{eqn2}) and
the number of $\cos$-terms
in $\phi(k)$ (in this case one term) complies
with the estimate (\ref{estimate}).
Because of
$|r_2|<1$, it is the spectral equation of a
regular quantum graph.
Using the explicit form (\ref{weight1}) of
the coefficients $A_{p}$ in
the expansion (\ref{kn}), we obtain the explicit
series expansion for
every root $k_n$ of (\ref{3hydra}).
Thus the spectrum of the scaling step potential
shown in Fig.~3 may be calculated explicitly
and analytically with the help of (\ref{kn}).
This by itself
is a considerable advance in the theory of
simple one-dimensional quantum systems which
up to now could only be solved using graphical
or numerical techniques \cite{LL,Flugge,Schiff}.

We illustrate the method and the convergence
of the series expansion
(\ref{kn}) with the following concrete, dimensioned
example of the scaling step potential
of Fig.~3. Choosing
$b=0.3$, $\lambda_1=0$,
and $\lambda_2=1/2$, we
computed the solutions
$k_1$, $k_{10}$ and $k_{100}$ of
(\ref{3hydra})
using three different
methods: (i) exact numerical,
(ii) explicit periodic-orbit expansion (\ref{kn}) of
the individual eigenvalues and
(iii) numerical integration using the
$S$-matrix representation (\ref{exp1})
of the density of states.
Addressing (i) we obtained the exact numerical
values of the
three selected roots
of (\ref{3hydra}). The result is:
$k_1=4.107149$,
$k_{10}=39.305209$ and $k_{100}=394.964713$.

Turning to method (ii) we re-computed
these three eigenvalues using (\ref{kn}) directly
including progressively more periodic orbits in the
expansion (\ref{kn}). The result is presented in
Table~1 which shows the values of $k_1$, $k_{10}$ and
$k_{100}$ computed with (\ref{kn})
including periodic orbits coded by binary words
of length $l=5$, 10, 15 and 20, respectively. This corresponds
to including 23, 261, 4807, and 111321 periodic orbits in
the expansion (\ref{kn}), respectively. We observe that
the accuracy does not improve monotonically, but that
there is a definite overall improvement of accuracy with
the code length. As a matter of fact,
as discussed above and shown mathematically in
\cite{JMP}, the series (\ref{kn}) converges, and converges
to the exact results of $k_n$ in the
limit of $l\rightarrow\infty$.

Turning to method (iii) we note that due to the exponential
proliferation of periodic orbits, it becomes progressively
more difficult to compute the codes of longer
periodic orbits.
Nevertheless, with the help of a numerical procedure,
we are able to illustrate the convergence behavior
of (\ref{kn}) for code lengths much longer than $l=20$.
Starting from (\ref{exp1}) we
compute the $S$-matrix numerically and perform all the steps
leading up to (\ref{kn}) numerically.
In particular this method involves numerical computation
of $S$-matrix powers and numerical integration over $k$.
Within any given level of numerical accuracy
this method is completely
equivalent to the method of summing the orbits, but allows us
to extend the computations such that we
effectively include all periodic orbits up to code length
$l=150$. This
corresponds roughly to $2^{150}=1.4\times 10^{45}$
periodic orbits, since
the periodic orbits on the three-vertex linear
graph are coded by a
binary code.
This estimate is substantiated by an analytical estimate
of the number of periodic orbits. For the three-vertex
graph the periodic orbits are binary necklaces over
the two symbols ${\cal L}$ and ${\cal R}$ \cite{JMP}.
The number of binary necklaces of length $\ell$
is given by \cite{combi}
\begin{equation}
{\cal N}(\ell) = {1\over \ell}\, \sum_{n|\ell}\,
\phi(n)\, 2^{\ell/n},
\label{ncount}
\end{equation}
where the symbol ``$n|\ell$'' denotes ``$n$ is a divisor
of $\ell$'', and
$\phi(n)$ is Euler's totient function defined as the
number of positive integers smaller than $n$ and relatively
prime to $n$ with $\phi(1)=1$ as a useful convention.
An upper limit for $\Lambda$ is obtained if we use
(\ref{ncount}) in the case where $\ell$ is a prime
number. In this case (\ref{ncount}) reduces to
\begin{equation}
{\cal N}(p)={1\over p}\,
\left[\phi(1)\, 2^p + \phi(p)\, 2^1\right]=
{1\over p}\left[2^p+2(p-1)\right],
\label{pcount}
\end{equation}
where $p$ is prime. Thus, in the limit of $p\rightarrow
\infty$ we have ${\cal N}(p)\rightarrow 2^p/p$, and
therefore, as an upper limit
\begin{equation}
\Lambda =
\lim_{p\rightarrow\infty}
{\ln\left[\#(p)\right]\over p}\ \leq\
\lim_{p\rightarrow\infty}
{\ln\left[p{\cal N}(p)\right]\over p}=\ln(2).
\label{esti}
\end{equation}
Thus, according to this estimate, the total number of
periodic orbits of length 150 is again
\begin{equation}
\#(150)\sim e^{150\Lambda}=2^{150}.
\end{equation}
We also computed numerical estimates of $\Lambda$.
Using the exact formula (\ref{ncount}) for counting
periodic orbits in (\ref{topo}) and
including periodic orbits with code lengths of up
to $l=1000$, we found $\Lambda>1.987$,
consistent with the estimate (\ref{esti}). For $l=150$,
relevant for our numerical example,
the asymptotic regime is not yet reached and
we find $\Lambda\approx \ln(1.943)$.
This value for $\Lambda$ can be used for a more refined
estimate of the number of periodic orbits of length
$l=150$, $\#(150)\approx 1.943^{150}\approx 2\times 10^{43}$.
Clearly, computing the codes of that many
periodic orbits and summing them up
in (\ref{kn})
is beyond the storage capacity and power
of any currently existing computer,
but is apparently
no obstacle to the numerical simulation of
that many periodic orbits included in (\ref{kn}). Figure~4
illustrates the rate of
convergence of the eigenvalues
$k_1$, $k_{10}$ and $k_{100}$ obtained with method (iii) as
a function of code length $l$ up to $l=150$.
Shown is the relative error
$\epsilon_l=|k_n(l)-k_n|/k_n$ for $n=1$, 10 and 100.
The error is seen to decrease on average as a function of
increasing periodic-orbit length $l$. From Fig.~4
we obtain approximately
$\epsilon_l=|k_n(l)-k_n|/k_n\sim 1/l^2$ on average.

 For a four-vertex linear graph, the spectral equation is
\begin{eqnarray}
\sin(S_0^0 k)=r_3 \sin(\Omega_1 k)-
r_2 r_3 \sin(\Omega_2 k) +
r_2\sin(\Omega_3 k),
\label{4vertex}
\end{eqnarray}
where
$S_0^0=S_1^0+S_2^0+S_3^0$,
$\Omega_1=S_1^0+S_2^0-S_3^0$,
$\Omega_2=S_1^0-S_2^0+S_3^0$,
$\Omega_3=S_1^0-S_2^0-S_3^0$
and $r_2$, $r_3$
are the reflection coefficients
at the vertices $V_2$ and $V_3$, respectively.
With $\gamma_i=\pi/2$, $i=0,1,2,3$,
this spectral equation is of the form
(\ref{eqn1}), (\ref{eqn2}) and the number of
$\cos$-terms in $\phi(k)$ (three in this case)
complies with (\ref{estimate}). For
\begin{equation}
|r_3|+|r_2r_3|+|r_2|<1,
\label{max}
\end{equation}
the four-vertex linear graph is regular.
In this case the energy values of the four-vertex linear
graph may be calculated exactly using the periodic-orbit
expansion (\ref{kn}). According to Fig.~5 the
set of $r_2$, $r_3$ values that fulfill (\ref{max})
occupies a diamond-shaped area bounded by the
functions $r_3=\pm(1-|r_2|)/(1+|r_2|)$. This observation
proves that regular quantum graphs are an important, finite-measure
subset of quantum graphs.

The set of regular quantum graphs is much wider than
the three- and four-vertex quantum graphs discussed
in detail above. Since, as indicated by (\ref{3hydra}) and
(\ref{4vertex}), the amplitudes $a_i$ in
(\ref{eqn2}) involve products of vertex reflection
coefficients, and since the vertex
reflection coefficients of a linear quantum graph
(via proper choice of the bond potentials)
are free parameters of the quantum graph, a
finite-measure set of regular quantum graphs
exists for {\it any} given linear graph.
It is possible that more complex graph topologies,
such as rings and stars, may also
admit a set of regular quantum graphs.
This topic is currently under investigation.

\section{Discussion, Summary and Conclusion}
In this paper we defined and studied a subset of
quantum graphs: regular quantum graphs. Regular
quantum graphs satisfy the regularity condition
(\ref{reg}) which implies that the roots of the
spectral equation are
confined to regularly spaced root intervals.
One and only one root is found per root interval.
This property allows us to
derive rigorous, converging periodic-orbit expansions
for individual energy levels of regular quantum graphs.

We often
hear the comment that the expansion (\ref{kn}) for
$k_n$ cannot possibly converge, since (\ref{ro})
is divergent. This comment is invalid. There is
a fundamental difference between (\ref{ro}) and
(\ref{kn}). Equation (\ref{ro}) is a
periodic-orbit expansion of the kernel
of a functional (a series of Dirac
delta ``functions''),
whereas (\ref{kn}) is a
periodic-orbit expansion of a simple c-number.
On the level of (\ref{ro}) the concepts
of convergence or divergence are not even
defined. Only after multiplying (\ref{ro})
with a test function (a function with
compact support \cite{distri}) and integrating
over $k$, are the concepts of divergence
and convergence defined. In this sense even
(\ref{ro}) is convergent. This is also known
as convergence in the distribution sense
\cite{distri} and leads to proper convergence
in the usual sense of elementary
undergraduate-level analysis
after multiplying with the test function
and integrating. Thus (\ref{kn}) converges
in the usual sense of elementary calculus since
our method of multiplying (\ref{kn}) with
$k$ and integrating from $\hat k_{n-1}$ to
$\hat k_n$ is equivalent with multiplying
(\ref{ro}) with the test function
\begin{equation}
g(k)=\cases{1, &for $\hat k_{n-1}\leq k\leq \hat k_n$,\cr
            0, &otherwise,\cr }
\end{equation}
and integrating over $k$. Since $g(k)$ is
a proper test function \cite{distri}, the
convergence of (\ref{kn}) is no longer
a mystery.

Another more serious comment concerns the sense in which
quantum graphs are classically
chaotic. Quantum graphs are based
on a one-dimensional network of vertices and bonds.
Therefore a dynamical Liapunov exponent \cite{Gutzw,Ott}
cannot be defined in the traditional sense of
exponentially diverging initially close trajectories.
As explained in \cite{QGT1}, however, this is no
obstacle to associate a classical phase space with
a quantum graph and to show that the classical
dynamics in this phase space is mixing \cite{QGT1}.
Moreover, quantum graphs fulfill another property
of quantum chaos, the exponential proliferation of
classical periodic orbits as manifested by a
positive topological entropy (see Sects. III, IV).
Therefore, quantum graphs have been called
``paradigms of quantum chaos'' \cite{QGT5}.
Inasmuch as the positive topological entropy is concerned
regular quantum graphs qualify as quantum chaotic
systems. Regular quantum graphs do not show all
the characteristics of ``fully developed''
quantum chaos. For instance,
because of the existence of the
forbidden zones $R_n$ (see Sect. III) they definitely
do not show a Wignerian nearest neighbor statistics.
We do not believe that this is a problem
since there is no universally accepted
rigorous definition of quantum chaos that
requires ``Wignerian statistics'' as one of the
necessary conditions.
The only broadly accepted criterion is that
``quantum chaos'' deals with quantum systems that
are chaotic in their classical limit. Based on
this criterion, together with the mixing property \cite{QGT1}
and the positive
topological entropy (see Sect. IV) regular
quantum graphs definitely qualify as quantum
chaotic systems, i.e. systems chaotic in the classical limit.

A more delicate point concerns the derivation of
(\ref{kn}) by integrating the fluctuating part in
(\ref{ro}) term-by-term. Since the resulting
periodic-orbit expansion is only conditionally convergent,
this is reason for concern. We addressed this point
from a rigorous mathematical point of view in
\cite{JMP}. We were able to prove rigorously that
the interchange of integration and summation in
the $k$ integral of $k\rho(k)$ is allowed. Thus,
term-by-term integration of (\ref{ro}) is
justified, validating the final result (\ref{kn}).

To our knowledge this is the
first time that the
energy levels of a class of
classically chaotic systems
are expressed one by one with the help of convergent
periodic-orbit expansions.
Studying specific examples of regular quantum graphs
we proved that the class of regular quantum graphs is
not empty; that it is in fact an important
finite-measure subset
of quantum graphs.
The explicit formulae of individual quantum energy levels
obtained in this paper remind us of
the Einstein-Brillouin-Keller (EBK) method \cite{Gutzw}
for the
quantization of integrable classical systems.
But there are important differences. Regular quantum
graphs do not correspond to classically integrable
systems. In fact, due to the importance of non-Newtonian
periodic orbits, the number of classical periodic orbits
proliferates exponentially with the code-length. This
proves that even regular quantum graphs, as defined by
the regularity condition (\ref{reg}), are classically
chaotic systems with a positive topological entropy. For
one of our examples, the scaling step potential
discussed in Sect. IV,
we computed estimates of
the topological entropy analytically and numerically.
In both cases
it turned out to be positive and close to $\ln(2)$.
This proves
that, at least for the cases studied,
the classical limit is chaotic.
Another difference to EBK theory is that the periodic
orbits in regular quantum graphs are not confined to
phase-space tori. Finally, in contrast with
EBK theory,
a semiclassical theory which does not usually return
exact results, our formulae are mathematically exact.
In summary, despite the apparent complexity
and exponential proliferation of
the periodic orbits of regular quantum graphs,
the organization of the roots of the spectral equation
into regularly spaced intervals makes it
possible to pinpoint every single energy eigenvalue
of a regular quantum graph analytically and
exactly by an
explicit, convergent periodic-orbit expansion.

Y.D. and R.B. gratefully acknowledge financial
support by NSF grants
PHY-9900730 and PHY-9984075; Y.D. and R.V.J. by NSF
grant PHY-9900746.

%%%%%%%%%%%%%%%%%%%%%%%%%%%%%%%%%%%%%%%%%%%%%%%%%%%%%%%%%%%%%%%%

\pagebreak

\centerline{\bf Figure Captions}

\bigskip \noindent
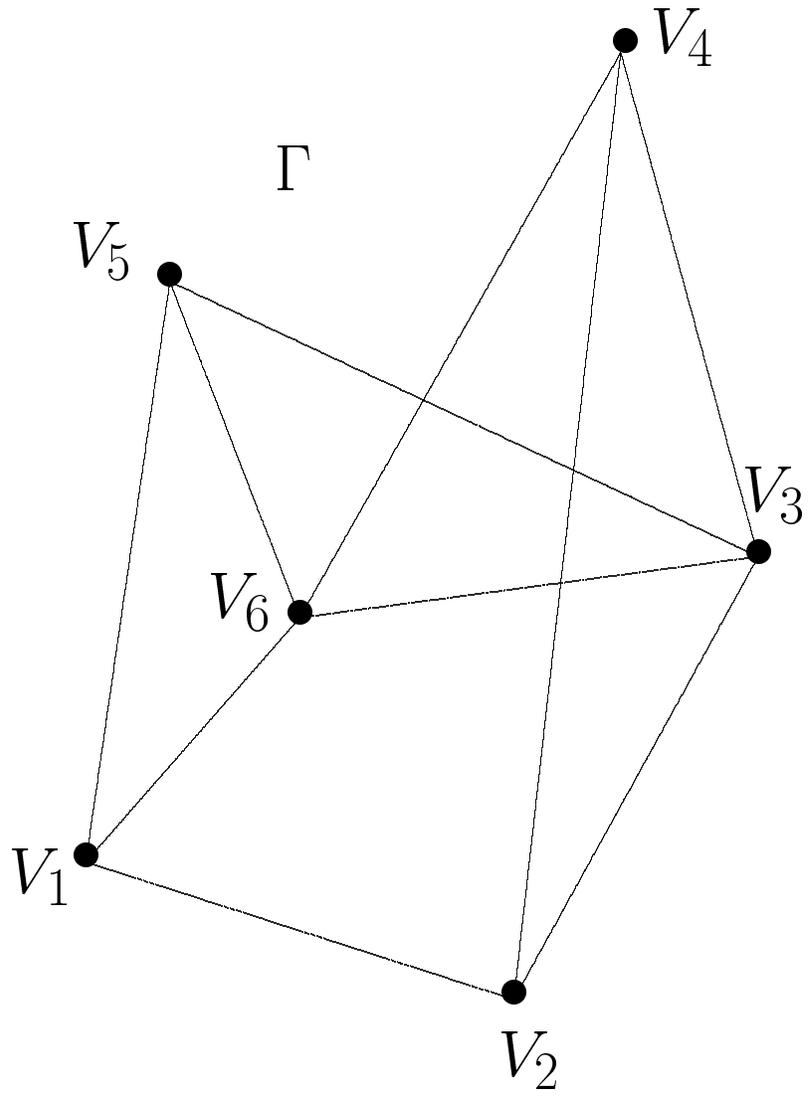
{\bf Fig.~1:} A generic non-planar
              graph with six
              vertices and ten bonds.

\bigskip \noindent
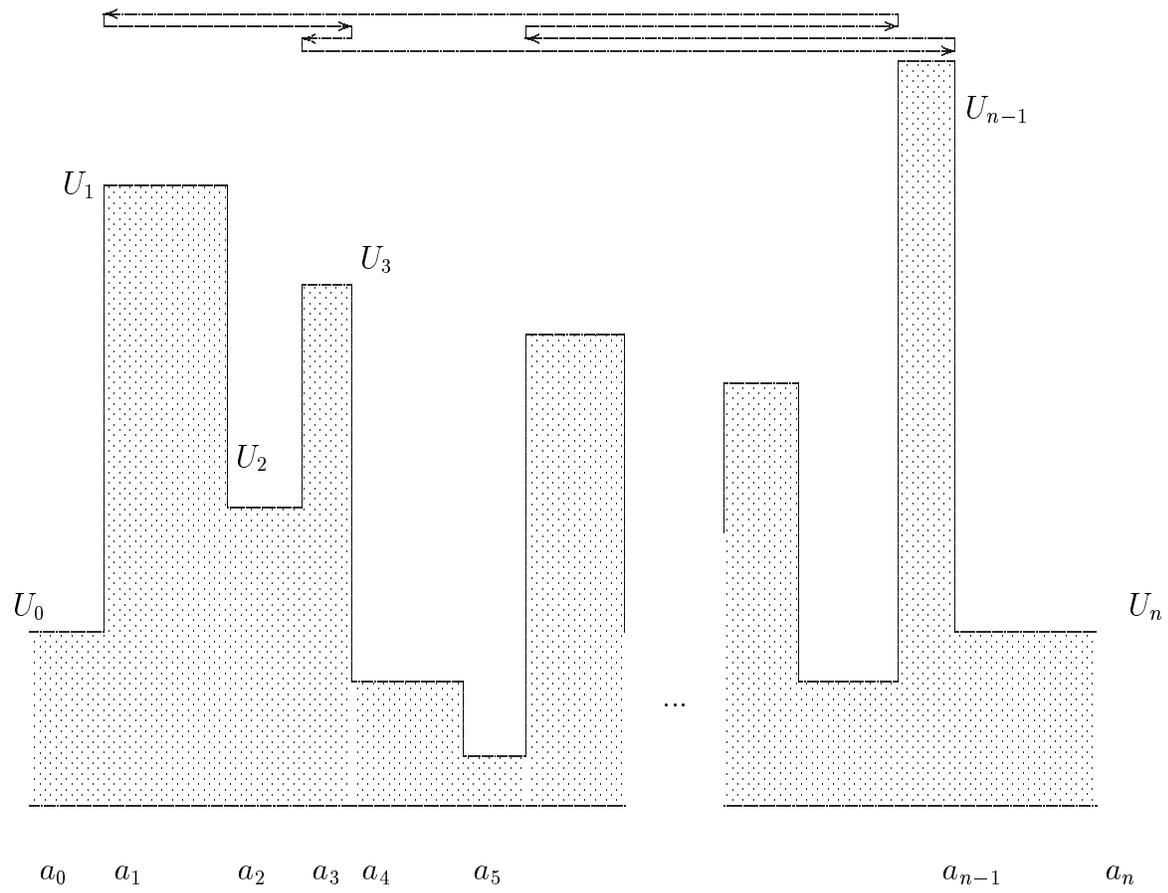
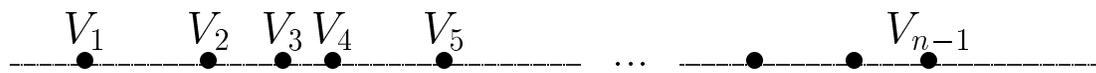
{\bf Fig.~2:} An example of a (Manhattan) step
              potential (a) and its associated
              linear graph (b).
              Also shown is a
              non-Newtonian periodic orbit characterized
              by six above-barrier reflections.

\bigskip \noindent
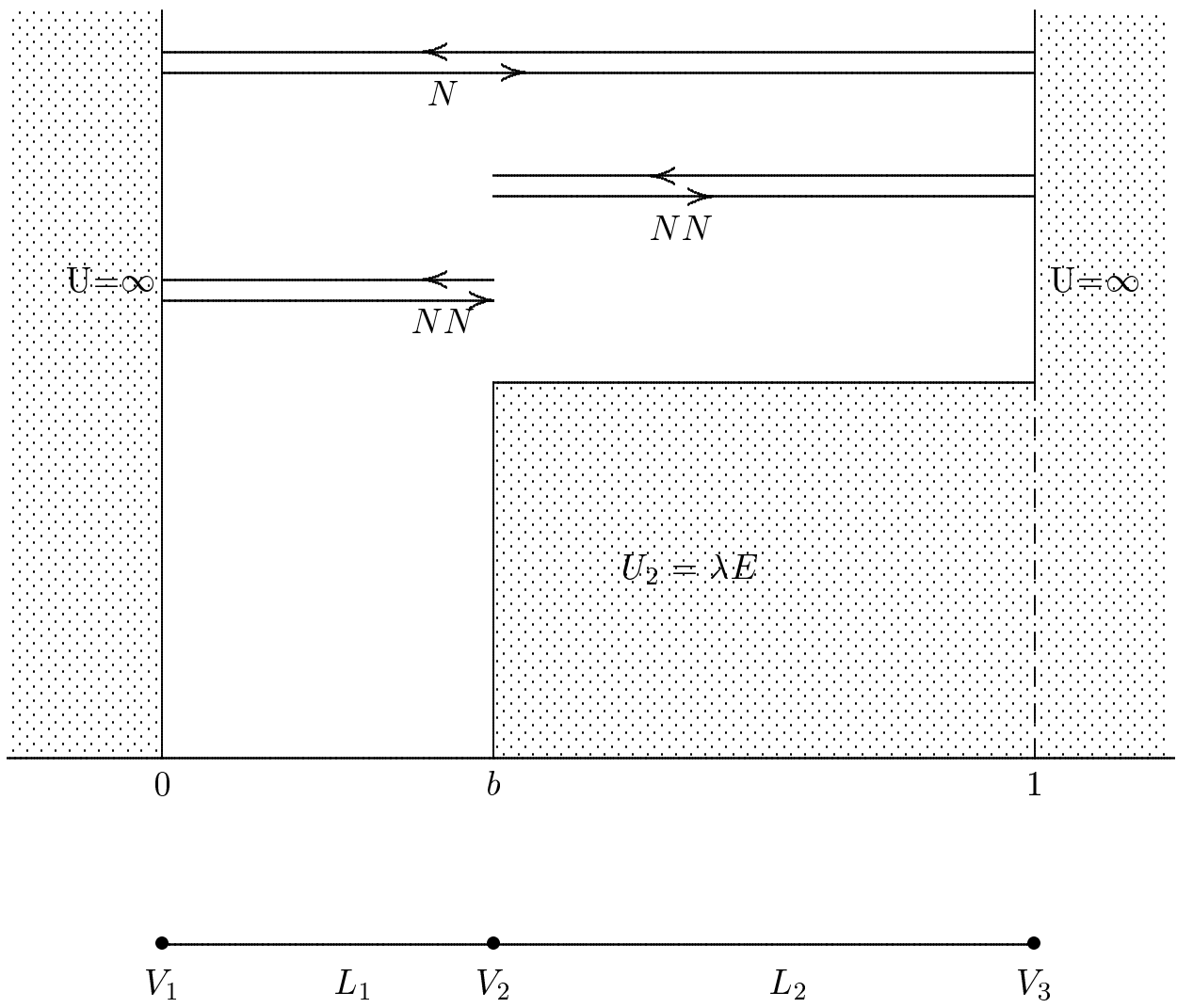
{\bf Fig.~3:} A scaling step potential (top),
              equivalent to a
              three-vertex linear graph (bottom), as an
              example of a regular quantum graph. A
              Newtonian ($\cal{LR}$) and two
              non-Newtonian ($\cal{L}$, $\cal{LRR}$)
              periodic orbits are also shown
              together with their $\cal{L-R}$ codes.

\bigskip \noindent
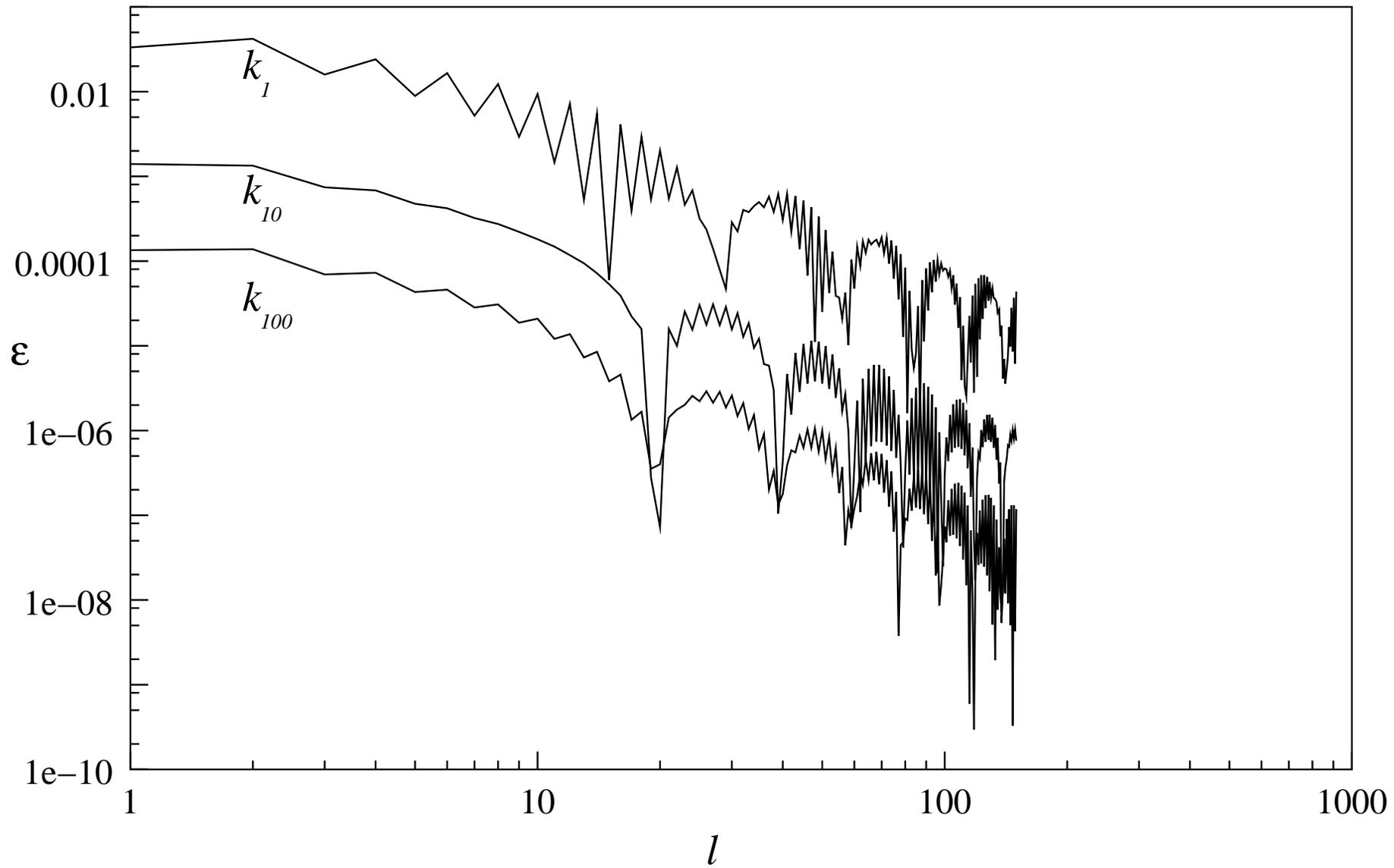
{\bf Fig.~4:} The deviation
              $\epsilon_l=|k_n(l)-k_n|/k_n$
              of the exact eigenvalues
              for $k_{1}$, $k_{10}$, and $k_{100}$
              from the corresponding values obtained
              via the series representation,
              as a function of the lengths
              $l$ of the periodic orbits.

\bigskip \noindent
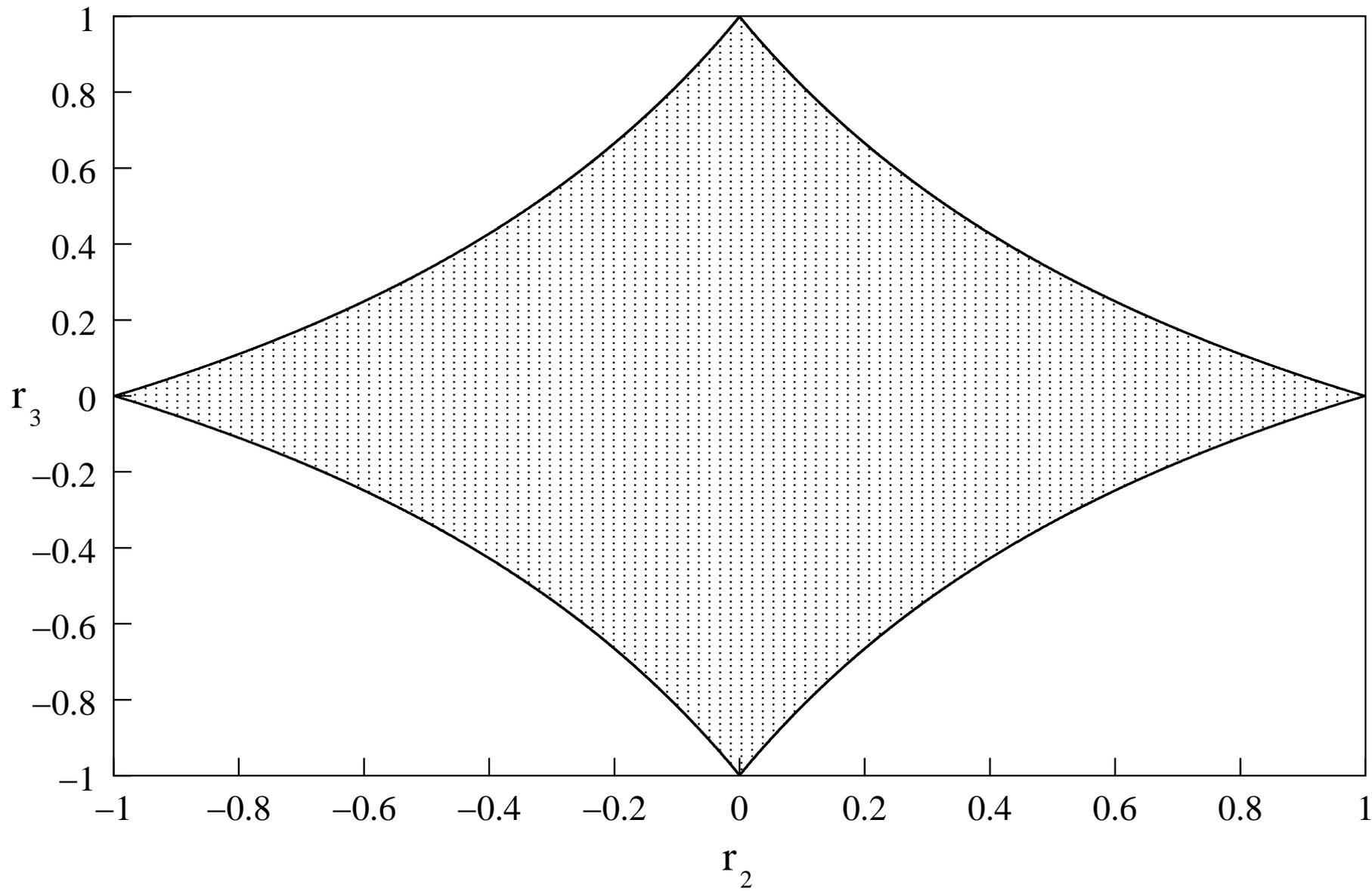
{\bf Fig.~5:} Parameter space $(r_2,r_3)$ of
              a four-vertex linear quantum graph.
              Parameter combinations in the shaded
              region correspond to regular quantum graphs.
              This demonstrates that the subset of
              regular quantum graphs within the set of
              all four-vertex linear quantum graphs
              is non-empty and in fact of finite measure.

\vskip1cm
\begin{table}[htbp]
                    \begin{center}
                      \leavevmode
                      \begin{tabular}[h]{|c|c|c|c|c|c|c|}
\hline
             root      & $l=5$   &  $l=10$ &  $l=15$   &  $l=20$
             &   exact   & error   \\
\hline
             $k_{1}$   &   4.11608 & 4.11653   & 4.10721   & 4.10513
             & 4.10715   & 0.00202 \\
             $k_{10}$  &  39.28658 & 39.29807  & 39.30730  & 39.30521
             & 39.30521  & 0.00000 \\
             $k_{100}$ & 394.94770 & 394.95647 & 394.96622 & 394.96456
             & 394.96471 & 0.00016 \\
\hline
                      \end{tabular}
                    \end{center}
                  \end{table}

\vskip1cm
\centerline{\bf Table Captions}
\bigskip\noindent
{\bf Table~1:} Successive approximations of the
eigenvalues $k_1$ (first row), $k_{10}$ (second row) and
$k_{100}$ (third row) of a specific scaling step-potential
(see text for details)
as a function of code length $l$ (columns
2--5).
The exact values of $k_1$, $k_{10}$ and $k_{100}$
are listed in column 6. Column 7 lists the absolute errors
$|k_n(l=20)-k_n|$ for $n=1$, 10 and 100.

\begin{thebibliography}{99}

\bibitem{Gutzw}  M. Gutzwiller, {\it Chaos in
                 Classical and Quantum Mechanics}
                 (Springer, New York, 1990).

\bibitem{Opus}  Yu. Dabaghian, R. V. Jensen, and R. Bl\"{u}mel,
                Pis'ma v ZhETF {\bf 74}, 258 (2001);
                JETP Lett. {\bf 74}, 235 (2001).

\bibitem{QGT1}  T. Kottos and U. Smilansky, Phys. Rev.
                Lett. {\bf 79}, 4794 (1997);
                Ann. Phys. (N.Y.) {\bf 274}, 76
                (1999).

\bibitem{QGT2} E. Akkermans, A. Comtet,
                J. Desbois, G. Montambaux, and
                C. Texier, {\it Spectral determinant
                on quantum graphs}
                (LANL archive cond-mat/9911183, 1999).

\bibitem{QGT3}  M. Pascaud and
                G. Montambaux, Phys. Rev. Lett. {\bf 82},
                4512 (1999).

\bibitem{QGT4} F. Barra and P. Gaspard, Phys. Rev. E
                  {\bf 63}, 066215 (2001).

\bibitem{QGT5} U. Smilansky, J. Phys. A (Math. and Gen.)
                   {\bf 33}, 2299 (2000).

\bibitem{BGS}  O. Bohigas, M.-J. Giannoni, and
               C. Schmidt, Phys. Rev. Lett.
               {\bf 52}, 1 (1984).

\bibitem{LH}  M.-J. Giannoni, A. Voros, and J. Zinn-Justin,
              {\it Chaos et Physique
              Quantique-Chaos and Quantum Physics},
              Les Houches session LII, 1989
              (Elsevier Science Publishers, Amsterdam,
              1991).

\bibitem{Couch} L. Couchman, E. Ott, and T. M. Antonsen, Jr.,
                Phys. Rev. A {\bf 46}, 6193 (1992).

\bibitem{RS1}  R. E. Prange, E. Ott,
               T. M. Antonsen, B. Georgeot, and R.
               Bl\"{u}mel, Phys. Rev. E {\bf 53}, 207 (1996).

\bibitem{RS2}  R. Bl\"{u}mel,
               T. M. Antonsen, Jr., B. Georgeot, E. Ott, and
               R. E. Prange, Phys. Rev. Lett. {\bf 76}, 2476 (1996);
               Phys. Rev. E {\bf 53}, 3284 (1996).

\bibitem{Pastur} I. M. Lifshits, S. A. Gredeskul,
                 and  L. A. Pastur,
                {\it Introduction to the theory of
                disordered systems}
                (Wiley Interscience, New York, 1988).

\bibitem{LL} L. D. Landau and E. M. Lifshitz,
             {\it Quantum Mechanics} (Pergamon, Oxford, 1960).

\bibitem{RS3} L. Sirko, P. M. Koch, and R. Bl\"umel,
               Phys. Rev. Lett. {\bf 78}, 2940 (1997).

\bibitem{Bauch} Sz.\ Bauch, A. B{\l}\c{e}dowski, L. Sirko,
               P. M. Koch, and R. Bl\"umel,
               Phys.  Rev.  E {\bf 57}, 304 (1998).

\bibitem{Nova} Y. Dabaghian, R. V. Jensen, and
               R. Bl\"umel, Phys. Rev. E {\bf 63}, 066201 (2001).

\bibitem{JMP} R. Bl\"umel, Yu. Dabaghian, and R. V. Jensen,
              {\it Mathematical Foundations of
              Regular Quantum Graphs}, in
              preparation for submittal to the J. Math. Phys. (2001).

\bibitem{Forma} R. Bl\"umel and Yu. Dabaghian,
                {\em Combinatorial identities
                for binary necklaces
                from exact ray-splitting
                trace formulae}, J. Math. Phys.,
                in press (2001).

\bibitem{Flugge} S. Fl\"ugge,
                {\it Practical Quantum Mechanics I} (Springer,
                New York, 1971), Problem 26.

\bibitem{Schiff} L. I. Schiff, {\it Quantum Mechanics},
                 (McGraw-Hill, New York, 1955).

\bibitem{combi} J. Riordan, {\it An Introduction to Combinatorial
                Analysis} (John Wiley \& Sons, New York, 1958).

\bibitem{distri} W. Walter, {\it Einf\"uhrung in die Theorie
                 der Distributionen} (Bibliographisches Institut,
                 Mannheim, 1974). %@

\bibitem{Ott} E. Ott, {\it Chaos in Dynamical Systems}
              (Cambridge University Press, Cambridge, 1993). %@

\end{thebibliography}
\end{document}